\documentclass[11pt,aps,amsmath,amssymb,preprint,nofootinbib]{revtex4}
\usepackage[active]{srcltx}
\usepackage[utf8]{inputenc}
\usepackage{latexsym}
\usepackage{amsmath}
\usepackage{graphicx}
\usepackage{slashed}
\usepackage{xcolor}

\begin{document}

\title{Nonrelativistic Limit of Dirac Theory From Effective Field Theory}

\author{Rodrigo Corso B. Santos}
\email{rodrigocorso@uel.br}
\affiliation{Departamento de F\'isica, Universidade Estadual de Londrina, 
Caixa Postal 10011, 86057-970, Londrina, PR, Brasil}

\author{Pedro R. S. Gomes}
\email{pedrogomes@uel.br}
\affiliation{Departamento de F\'isica, Universidade Estadual de Londrina, 
Caixa Postal 10011, 86057-970, Londrina, PR, Brasil}

\begin{abstract}

In this work we analyze the low-energy nonrelativistic limit of Dirac theory in the framework of effective field theory. By integrating out the high-energy modes of Dirac field, given in terms of a combination of the two-components Weyl spinors, we obtain a low-energy effective action for the remaining components, whose equation of motion can then be compared to the Pauli-Schr{\"o}dinger equation after demanding normalization of the wave function. We then discuss the relevance of the terms in the effective action in the context of an anisotropic dimensional analysis which is suitable for nonrelativistic theories.

\end{abstract}

\maketitle


\section{Introduction}\label{S1}

Laws of physics embody two remarkable properties: {\it scale dependence} and {\it decoupling} \cite{Gross,Gomes1}. Scale dependence means that the physics we observe depends on the scale we are doing so. In field theory problems we usually refer to a length scale or, equivalently, an energy scale. The macroscopic physics characterizing, say, the thermodynamics of a fluid is very different of the quantum mechanical properties of particles that constitute it. The decoupling is the property that enables us to describe the macroscopic behavior of the fluid according to thermodynamics without knowing about the quantum mechanics of atoms. The important observation is that at each scale, distinct degrees of freedom are relevant to characterize the system. 

Effective field theory is the suitable framework to take advantage of those two properties, providing a systematic method for isolating the relevant low-energy physics from high-energy physics \cite{Polchinski,Georgi,Kaplan,Petrov}. We then consider independent dynamics for such degrees of freedom encoded in a low-energy effective action. This is not always a straightforward task. Two helpful ingredients in this analysis are symmetries and dimensional analysis.

We usually classify effective field theories according to the way they are constructed: {\it top down} as opposed to {\it bottom up} \cite{Georgi}. In the top down approach, we know the higher energy theory and obtain the low-energy one by eliminating (integrating out) the high-energy modes, as we will review below. In this case, the low-energy theory is not only a convenience but it also makes easier to unveil the low-energy physical properties. In the bottom up approach, we do not know the microscopic mechanisms of the system or it is hard to get information from the microscopic equations. In this case, the effective action is constructed according to symmetries and the dimensional analysis often enables us to consider only a few leading terms, which are responsible for the low-energy dynamics.  

In this work we consider the low-energy nonrelativistic limit of the Dirac theory. The traditional way to explore this regime is based essentially on the identification of the components of the Dirac spinor that are important at low energies compared to the electron mass, and then isolating them through the use of the Dirac equation (see, for example, \cite{Bjorken,Gottfried,Costella}). A powerful method to implement this in a systematic way is known for a long time and is due to Foldy and Wouthuysen \cite{Foldy}, which corresponds to a canonical transformation that naturally decouples the low-energy components of the Dirac spinor, while keeping unitarity manifestly. It also provides a systematic way to access any given order in an expansion in the inverse of mass of the electron that is the relevant expansion parameter in this case. 

This paper follows that general spirit but in the context of effective field theories, which has the benefit of providing us with the dimensional analysis that is hidden in the usual quantum mechanical treatment. A related discussion can be found in \cite{Holstein} (for a discussion of nonrelativistic limit in lower dimensional field theory, see \cite{Silva}).
We start with the partition function defining the Dirac theory coupled to a background gauge field and then proceed to integrate out the high-energy modes of Dirac field, given in terms of a combination of the two-components Weyl spinors. Once we eliminate this "piece" of the Dirac spinor we end up with a low-energy effective action for the remaining components, whose equation of motion can then be compared to the Pauli-Schr{\"o}dinger equation after demanding normalization of the wave function. We then discuss the relevance of the terms in the effective action in the context of an anisotropic dimensional analysis which is appropriate for nonrelativistic theories. The whole analysis places the problem of investigating the nonrelativistic regime of Dirac theory in a more general context, which represents our current understanding of quantum field theories.   

This article is aimed at advanced graduate students as well as general physicists who do not necessarily work on the subject. The level of discussions assumes previous contact with some material that is typically covered in the core graduate courses on quantum mechanics and electrodynamics, as path integral quantization, relativistic quantum mechanics and the covariant formulation of electromagnetism. Some knowledge of field theory is welcome but not mandatory to understand the main points of the paper. We believe therefore that the present paper offers a complementary study to the traditional presentations. 

The work is organized as follows. In Sec. \ref{S2}, we review the general strategy for obtaining the effective action upon integration of high-energy degrees of freedom. Sec. \ref{S3} is devoted to the application of effective field theory methods to analyze the low-energy nonrelativistic regime of Dirac theory. We discuss the relevance of the operators in the effective action which are responsible for the correction of the $g$-factor of the electron, the anomalous correction due to the Pauli coupling and also for the gross and fine structure of the electron in the presence of a central static electric field. We conclude with a summary and additional remarks in Sec. \ref{S4}.  Subsidiary calculations are carried out in three appendices.

\section{Effective Field Theory}\label{S2}

In this section we review some important elements involved in the effective field theory approach \cite{Polchinski,Georgi,Kaplan,Petrov}. Suppose we have a theory with a characteristic energy scale $E_0$ and we are interested in the physics in a scale $E\ll E_0$. Let the theory be given in terms of a field $\phi$ (or a set of fields) and be defined by the partition function in the presence of a background field $A$:
\begin{equation}
Z[A]=\int\mathcal{D}\phi e^{i S[\phi,A]},
\label{1.1}
\end{equation}
where $S[\phi,A]=\int d^Dx\mathcal{L}(\phi,A)$ and $D$ is the spacetime dimensionality.  

We consider an ultraviolet cutoff $\Lambda_{0}\sim E_0$ and we split the field in low and high-energy modes compared to the cutoff, $\phi=\phi_L+\phi_H$. This seems a little vague but it can be achieved, for example, in terms of the Fourier decomposition of the field
\begin{equation}
\phi(x) =\underbrace{\int_{|k|<\Lambda_{0}}\frac{d^Dk}{(2\pi)^D}e^{-ikx}\phi(k)}_{\phi_L}+\underbrace{\int_{|k|>\Lambda_{0}}\frac{d^Dk}{(2\pi)^D}e^{-ikx} \phi(k)}_{\phi_H}.
\label{1.2}
\end{equation}
The precise way this splitting is implemented is not important for us at this moment. We proceed by integrating out the high-energy field $\phi_H$
\begin{eqnarray}
Z[A]&=&\int \mathcal{D}\phi_H \mathcal{D}\phi_L e^{iS[\phi_H,\phi_L,A]}\nonumber\\
&=&\int \mathcal{D}\phi_Le^{iS_{eff}[\phi_L,A]},
\label{1.3}
\end{eqnarray}
where the low-energy effective action is formally defined as
\begin{equation}
e^{iS_{eff}[\phi_L,A]}\equiv \int \mathcal{D}\phi_H e^{iS[\phi_H,\phi_L,A]}.
\label{1.4}
\end{equation}

An important point is that we can expand the effective action $S_{eff}$ in terms of a set of local operators $O_a(x)$,
\begin{equation}
S_{eff}=\int d^Dx \sum_a \lambda_a O_{a}(x).
\label{1.5}
\end{equation}
We could worry about the locality of the effective action since as we are eliminating (integrating out) high-energy modes, this will precludes the observation of distance scales $\lesssim 1/\Lambda_0$, making it nonlocal in such scales. But as the effective action contains only low-energy fields, the corresponding typical distances involved are of magnitude $\gg 1/\Lambda_0$, and thus the low-energy effective action will be {\it local} in these scales. It means that the terms in the effective action involve fields and derivatives of fields that can be organized in a derivative expansion, which we represent schematically as $\partial/\Lambda_0$. We then expect that for the energies of interest, $E\ll \Lambda_0$, such terms will correspond to different powers of $E/\Lambda_0$. We will make this association a little more precise in a moment.

The local expansion (\ref{1.5}) is particularly useful in the case of bottom up approach, since we do not have access to the higher energy theory. The problem is that the action (\ref{1.5}) contains in principle an infinite number of terms. The only restriction so far is that the terms must in general be compatible with certain symmetries. To make it useful we need something else. Dimensional analysis gives the required element. It enables us to unveil the behavior of the terms of $S_{eff}$ and then identify the operators that will give the most important contributions at low energies. Let us consider the dimension of the operator $O_a$ in mass units as $[O_a]\equiv\Delta_a$. The dimension of the coupling constant is then $[\lambda_a]=D-\Delta_a$. Now we use the fact that the high-energy information (settled here by the cutoff $\Lambda_0$) is encoded in the low-energy theory through the parameters of the effective action, whereas the low-energy dynamics is dictated by the operator content. It means in practice that we can define dimensionless coupling constants incorporating the cutoff $\Lambda_0$, as $\tilde{\lambda}_a\equiv\Lambda_{0}^{\Delta_a-D} \lambda_a$, and for the contribution to the effective action of $\int d^Dx O_a(x)$, we estimate that for a process with a characteristic energy $E$,
\begin{equation}
\int d^Dx O_a(x)\sim E^{\Delta_a-D}.
\label{1.6}
\end{equation}
Thus the contribution of the $a$-th term to the effective action is
\begin{equation}
\tilde{\lambda}_a\left(\frac{E}{\Lambda_{0}}\right)^{\Delta_a-D}.
\label{1.7}
\end{equation}
As $E\ll \Lambda_{0}$, each operator in the effective action can have three distinct behaviors at low energies. If $\Delta_a-D>0$ the corresponding term is suppressed in the action. Such operators are called {\it irrelevant}. Terms with $\Delta_a-D<0$ grow at low energies and are called {\it relevant} operators. Operators with $\Delta_a-D=0$ are called {\it marginal} and give constant contribution to the effective action\footnote{The whole analysis of the relevance of operators in the effective action can be derived in a more systematic way. See, for example, Chap. 12 of Ref. \cite{Peskin}.}. The upshot of the analysis is that whenever we are concerned with the low-energy limit we have to consider essentially marginal and relevant operators in the effective action. Irrelevant operators will give only sub-leading corrections. In general this restricts the expansion in (\ref{1.5}) to contain a small number of terms.


\section{Low-Energy Limit of Dirac Theory}\label{S3}

Now we will follow the general strategy discussed previously to analyze the low-energy limit of the Dirac theory. The partition function of the Dirac theory minimally coupled to a background gauge field $A_{\mu}$ is
\begin{equation}
Z[A]=\int \mathcal{D}\bar{\Psi} \mathcal{D}\Psi \text{exp}~ i \int d^4x \bar{\Psi}(i \slashed{D}-m)\Psi,
\label{2.1}
\end{equation}
where $\slashed{D}\equiv \gamma^{\mu}D_{\mu}$, $D_{\mu}\equiv\partial_{\mu}+ieA_{\mu}$ and $\bar{\Psi}\equiv\Psi^{\dagger}\gamma^{0}$. In order to proceed with the identification of the low-energy part of the spinor field, we have to choose a representation for the Dirac matrices. Certainly, physics does not depend on this choice. We can start, for example, with the chiral representation,
\begin{align}
\gamma^{0}=\sigma^1\otimes \sigma^0=\left(\begin{array}{cc}
0 &  \textbf{I}_{2}\\ 
 \textbf{I}_{2} & 0
\end{array} \right) ~~~\text{and}~~~
\gamma^{i}=i\sigma^2\otimes\sigma^i=\left(
\begin{array}{cc}
0 & \textbf{$\sigma$}^{i} \\ 
-\textbf{$\sigma$}^{i} & 0
\end{array} \right),
\end{align}
where $\sigma^0\equiv\textbf{I}_{2}$ and $\sigma^i$ are the Pauli matrices. In this representation, the
Dirac spinor takes the form
\begin{align}
\Psi=\left(\begin{array}{c}
\varphi_{+} \\ 
\varphi_{-}
\end{array}\right), 
\label{psi dirac}
\end{align}
where $\varphi_+$ and $\varphi_-$ are two-component Weyl spinors. The Weyl spinors acquire a special status in the massless case. To see this, we write the free Dirac equation in the momentum space,
\begin{equation}
(\slashed{p}-m)\psi(p)=0.
\label{2.1aa}
\end{equation} 
In the massless case, it breaks up into two decoupled parts, 
\begin{equation}
\left(1+\frac{\vec{p}\cdot \vec{\sigma}}{|\vec{p}|}\right)\varphi_{+}(p)=0~~~\text{and}~~~\left(1-\frac{\vec{p}\cdot \vec{\sigma}}{|\vec{p}|}\right)\varphi_{-}(p)=0.
\label{2.1ab}
\end{equation}
Thus we see that the Weyl spinors are eigenstates of helicity, which is the projection of the spin along the momentum, $\vec{p}\cdot \vec{\sigma}/|\vec{p}|$. The mass term couples the spinors $\varphi_{+}$ and $\varphi_{-}$.

In terms of Weyl spinors the Dirac Lagrangian reads 
\begin{align}
\mathcal{L}=i\left(\varphi_{+}^{\dagger}D_{0}\varphi_{+}+\varphi_{-}^{\dagger}D_{0}\varphi_{-}-\varphi_{+}^{\dagger}\sigma^{i}D_{i}\varphi_{+}+\varphi_{-}^{\dagger}\sigma^{i}D_{i}\varphi_{-}\right)-m\left(\varphi_{-}^{\dagger}\varphi_{+}+\varphi_{+}^{\dagger}\varphi_{-}\right).
\end{align}
For convenience we define
\begin{equation}
 \Pi\equiv i \sigma^{k} D^{k}=\vec{\sigma}\cdot(-i\vec{\nabla}-e\vec{A}).
\end{equation}
To identify the low-energy components of the spinor field, we
consider the on-shell Weyl spinors in the momentum space $\varphi_{\pm}(\vec{p})$. In the rest frame we have $\varphi_{+}(\vec{p}=0)=\varphi_{-}(\vec{p}=0)$, as it follows directly from equation (\ref{2.1aa}), for $\vec{p}=0$.  In the presence of an electromagnetic field the particle cannot remains in rest but if the field is weak enough, it can be in a state where $|\vec{p}|\ll m$, such that $\varphi_{+}\sim \varphi_{-}$. Thus we expect that at the low-energy limit, the combination $\varphi_{+}-\varphi_{-}$ is suppressed by powers of $|\vec{p}|/m$. 

This invites us to introduce the low and high-energy components of the spinor as
\begin{align}
 \varphi_{L}\equiv\dfrac{1}{\sqrt{2}}\left(\varphi_{+}+\varphi_{-}\right)
 \label{2.2a}
\end{align}
and
\begin{align}
\varphi_{H}\equiv\dfrac{1}{\sqrt{2}}\left(\varphi_{+}-\varphi_{-}\right).
\label{2.2b}
\end{align} 
Incidentally, the change of basis from $\varphi_{\pm}$ to $\varphi_{R/L}$ corresponds to a different choice of representation for the Dirac matrices, namely, the Dirac representation, where the matrix $\gamma^0$ is diagonal (see Appendix A-2 of Ref. \cite{Zuber}). 
In terms of these new fields the Lagrangian becomes\footnote{As the transformations are linear they do not generate field-dependent factors in the functional Jacobian.}
 \begin{align}
 \mathcal{L}=\varphi_L^{\dagger}\left(iD_{0}-m\right) \varphi_L+\varphi_{H}^{\dagger}\left(iD_{0}+m\right) \varphi_H-\varphi_L^{\dagger}\Pi\varphi_H-\varphi_H^{\dagger}\Pi\varphi_L.
 \end{align}
In order to analyze the limit of low energies compared to $m$ it is convenient to extract a factor $e^{-imt}$ from the fields, so that we define the new fields $\varphi_L\equiv e^{-imt}\psi_L$ and $\varphi_H\equiv e^{-imt}\psi_H$. With this, the Lagrangian reads
 \begin{align}
  \mathcal{L}={\psi}_L^{\dagger}iD_{0} {\psi_L}+{\psi}_H^{\dagger}\left(iD_{0}+2m\right) {\psi}_H-{\psi}_L^{\dagger}\Pi{\psi}_H-{\psi}_H^{\dagger}\Pi{\psi_L}.
  \label{2.3}
 \end{align}
This Lagrangian makes it clear that the spinor $\psi_H$ plays the role a very massive field which is therefore suppressed at low energies compared to its mass $m$.
So now we are ready to integrate out the field $\psi_H$. This will produce the effective action $S_{eff}[\psi_L, A_0,\vec{A}]$, which will be local if we expand it in powers of inverse of the mass $m$.  It is also interesting to notice that as we are eliminating a "piece" of the Dirac spinor, Lorentz invariance is naturally lost in the effective action. That is the reason we wrote $A_0$ and $\vec{A}$ in $S_{eff}$ and not $A_{\mu}$.

As the Lagrangian is at most quadratic in $\psi_H$, we can straightforwardly to carry out the integration over it
 \begin{align}
 Z[A_0,\vec{A}]=\int \mathcal{D}\psi_L^{\dagger}\mathcal{D}\psi_L \exp i S_{eff}[\psi_L, A_0,\vec{A}],
 \label{2.4}
\end{align}
with the effective action given by
\begin{eqnarray}
S_{eff}[\psi_L,A_0,\vec{A}]&=&-i \text{Tr}\text{ln}(iD_0+2m)+\int d^{4} x\; \psi_L^{\dagger}(x)iD_{0}\psi_L(x)\nonumber\\&-&\int d^{4} x d^{4} y\;\psi_L^{\dagger}(y)\Pi(y) G(x,y)\Pi(x)\psi_L(x).
\label{2.5}
\end{eqnarray}
In this expression, $G(x,y)$ is defined as $G(x,y)\equiv \delta^{(3)}(\vec{x}-\vec{y})G_0(x,y)$, where $G_0(x,y)$ is the Green function of the operator $iD_0+2m$, 
\begin{equation}
(iD_{0}+2m)G_0(x,y)=\delta(x^0-y^0).
\label{2.5a}
\end{equation}
We are mostly interested in the static case, when the gauge fields do not depend on time, i.e., $A_0(x)=A_0(\vec{x})\equiv V(\vec{x})$ and $\vec{A}(x)=\vec{A}(\vec{x})$. In this case, the Green function becomes
\begin{equation}
G(x,y)=\dfrac{1}{2m}\sum\limits_{n=0}^{\infty}(-1)^{n}\left[\dfrac{i\partial_{x_{0}}-eV(\vec{x})}{2m}\right]^{n}\delta^{(4)}(x-y).
\label{2.6}
\end{equation}
The details of the computation of the effective action in (\ref{2.5}) and of the Green function in (\ref{2.6}) are discussed in Appendices \ref{AA} and \ref{AB}, respectively.

In our calculation we will consider the effective action for the first two terms in the above sum, i.e., $n=0,1$, which will give the contributions for the $g$-factor of the electron and for the fine structure. From now on we omit the index $L$ of $\psi_L$ and replace $\psi_L\rightarrow \psi$ for simplicity of notation whenever there is no risk of confusion. Thus it follows that
\begin{align}
S_{eff}&= \int d^4x \psi^{\dagger}i D_{0}\psi-\dfrac{1}{2m}\int d^{4}x \psi^{\dagger}\Pi^{2}\psi \nonumber\\
		&+\dfrac{1}{2m}\int d^{4}x d^{4}y \psi^{\dagger}(y)\Pi(y)\left(\dfrac{i\partial_{x_0}-eV(x)}{2m}\delta^{(4)}(x-y) \right) \Pi(x) \psi (x)  +\cdots\nonumber\\ 
		&=\int d^{4}x \left[\psi^{\dagger}i D_{0}\psi-\dfrac{1}{2m}\psi^{\dagger}\Pi^{2}\psi+\dfrac{1}{4m^{2}}\psi^{\dagger}\Pi (i\partial_{0}-eV)\Pi \psi \right]+\cdots.
		\label{2.7}
\end{align}
Note that, as the term $\text{Tr}\text{ln}(iD_0+2m)$ does not contribute to the equation of motion of $\psi$, we are just ignoring it. At this point is interesting to analyze the relevance of the terms in this action in light of dimensional analysis. Before doing this, let us consider the Dirac action in (\ref{2.1}). In the relativistic setting, where in mass units $[x^{\mu}]=-1$ and $[\Psi]=3/2$, the mass term is relevant, $[\bar{\Psi}\Psi]=3$. The dimension of the electromagnetic field is $[A_{\mu}]=1$. It means that the electromagnetic interaction $\bar{\Psi}\slashed{A}\Psi$ is marginal, i.e., $[\bar{\Psi}\slashed{A}\Psi]=4$. 

Now imagine we are going to lower and lower energies such that we need to classify the relevance of the terms of (\ref{2.7}). By using the relativistic dimensional setting for the action (\ref{2.7}), we have
\begin{eqnarray}
&[\psi^{\dagger}i D_{0}\psi]&=~4~~~\rightarrow~~~\text{marginal}\nonumber\\
&[\psi^{\dagger}\Pi^{2}\psi ]&=~5~~~\rightarrow~~~\text{irrelevant}\nonumber\\
&[\psi^{\dagger}\Pi D_0\Pi \psi]&=~6~~~\rightarrow~~~\text{irrelevant}.
\label{2.7a}
\end{eqnarray}
This is a little uncomfortable since the operator $\psi^{\dagger}\Pi^{2}\psi$ that contains the leading order spatial derivatives is irrelevant. It seems that at sufficiently low energies the field $\psi$ is no longer sensitive to spatial variations, which certainly does not correspond to reality. The point is that the relativistic attribution of dimensions is not adequate for the nonrelativistic action (\ref{2.7}).  Instead, it is more natural to consider an anisotropic setting, where the leading terms in time and spatial derivatives are treated on an equal foot, namely, we should impose that the operators $\psi^{\dagger} \partial_0 \psi$ and $\psi^{\dagger} \vec{\nabla}^2 \psi$ are of the same relevance in the effective action. This suggests that it is more suitable to consider a system of units where $[\partial_0]=[ \vec{\nabla}^2]$. We can use the mass scale $m$ to go to this new system of units.
This can be implemented precisely upon the rescaling $t\rightarrow mt$ and $A_0\rightarrow \frac{1}{m}A_0$\footnote{As the field $A_0$  has the same dimension as the time derivative, it should be rescaled accordingly. }, such that rescaled quantities will have dimensions
\begin{equation}
[t]=-2~~~\text{and}~~~[A_0]=2,
\label{2.7b}
\end{equation}
whereas the space-like quantities keep their dimensions, $[\vec{x}]=-1$ and $[\vec{A}]=1$. 
This is a typical behavior that occurs oftenly in condensed matter theories that are not Lorentz invariant. Such theories are named Lifshitz field theories \cite{Lifshitz} and are in general characterized by the so-called dynamical critical exponent $z$, which measures the degree of anisotropy between space and time,  $[t]=-z$ and 
$[\vec{x}]=-1$ (see also \cite{Fradkin,Horava1,Alexandre,Gomes2} for more recent developments). For the action (\ref{2.7}), it follows that $z=2$.

The rescaled action becomes
\begin{equation}
S_{eff}=\int d^{3}x dt \left[\psi^{\dagger}i D_{0}\psi-\dfrac{1}{2}\psi^{\dagger}\Pi^{2}\psi+\dfrac{1}{4m^{2}}\psi^{\dagger}\Pi (i\partial_{0}-eV)\Pi \psi \right]+\cdots.
\label{2.7c}
\end{equation}
In the anisotropic dimensional setting, we get\footnote{Notice that the dimension of the phase space integration is $[d^3x dt]=-5$ in anisotropic mass units.} 
\begin{eqnarray}
&[\psi^{\dagger}i D_{0}\psi]&=~5~~~\rightarrow~~~\text{marginal}\nonumber\\
&[\psi^{\dagger}\Pi^{2}\psi ]&=~5~~~\rightarrow~~~\text{marginal}\nonumber\\
&[\psi^{\dagger}\Pi D_0\Pi \psi]&=~7~~~\rightarrow~~~\text{irrelevant}.
\label{2.7d}
\end{eqnarray}
Now we see the point. The operator $\psi^{\dagger}\Pi^{2}\psi$ that was naively irrelevant in the analysis that led to (\ref{2.7a}) is actually marginal when we employ the appropriate dimensional analysis. This shows clearly the distinction between the terms of the action: the first two terms give the leading contribution (gross structure), whereas the third one gives a sub-leading contribution (fine structure).


\subsection{Comparison to the Pauli-Schr{\"o}dinger Equation}

It is a straightforward textbook exercise the comparison to the Pauli-Schr{\"o}dinger equation (see, for example, \cite{Gottfried}), but we discuss it here in the context of effective field theory. The equation of motion for the two-component spinor $\psi$ coming from the effective action (\ref{2.7}) reads
\begin{equation}
iD_{0}\psi=\dfrac{1}{2m} \Pi^{2}\psi-\dfrac{i}{4m^{2}}\Pi D_{0}\Pi\psi+\cdots.
\label{equacao diferencial psi dirac}
\end{equation}
It is tempting to compare this equation with the Pauli-Schr{\"o}dinger equation in order to extract the desired corrections. However, before a direct comparison  there is one point that we should be careful. Compliance with probabilistic interpretation shows that $\psi$ cannot be exactly identified with the Pauli-Schr{\"o}dinger wave function, denoted here by $\psi_s$, but only at leading order in the expansion of the inverse of the mass, i.e., $\psi=\psi_s+\mathcal{O}(1/m)$. If we go beyond the leading order, we must then amend this relation. Such corrections can be determined by considering the normalization condition that follows from the conserved current due to the global $U(1)$ symmetry of the Dirac Lagrangian 
\begin{align}
j^{\mu}=\bar{\Psi}\gamma^{\mu}\Psi.
\label{2.8}
\end{align}
In Appendix \ref{AC} we discuss how to construct this current via the Noether theorem.
By writing the normalized conserved charge $\int d^3x j^0$ in terms of two-component spinors $\psi_L$ and $\psi_H$, we obtain the normalization condition
\begin{align}
\int d^{3} x \left( \psi_L^{\dagger}\psi_L+\psi_H^{\dagger}\psi_H\right)=1, 
\label{normalizacao}
\end{align}
which shows that $\int d^3x \psi_L^{\dagger}\psi_L$ alone is not normalized. We have reinserted the $L$ and $H$ indexes in the spinors. 
The above expression should be identified with normalization condition of the Pauli-Schr{\"o}dinger wave function 
\begin{equation}
\int d^{3} x \psi^{\dagger}_{s}\psi_{s}=1.
\label{2.9}
\end{equation}
We must find $\psi_{s}$ as a function of $\psi_L$ that satisfies the equation (\ref{equacao diferencial psi dirac}). By taking the equation of motion of $\psi_H$ in the Lagrangian (\ref{2.3}), we get 
\begin{align}
\left(iD_{0}+2m\right)\psi_H-\Pi \psi_L=0.
\label{2.10}
\end{align}
Solving for $\psi_H$ in terms of the Green function we can write
\begin{equation}
\psi_H(x)=\int d^4y G(x,y) \Pi(y)\psi_L(y).
\label{2.11}
\end{equation}
The leading order terms of (\ref{2.6}) yield
\begin{align}
\psi_H=\dfrac{\Pi}{2m}\psi_L-\dfrac{iD_{0}\Pi}{4m^{2}}\psi_L+\cdots.
\label{relacao phi psi}
\end{align}
Using this result in (\ref{normalizacao}) and comparing with (\ref{2.9}), we obtain the required relation
\begin{align}
\psi=\left(1-\dfrac{\Pi^{2}}{8m^{2}}\right)\psi_{s}+\cdots. 
\label{relacao psis}
\end{align}
Finally, we substitute this relation in (\ref{equacao diferencial psi dirac}) to obtain
\begin{align}
iD_{0}\psi_{s}-\dfrac{1}{2m}\Pi^{2}\psi_{s}=\dfrac{i}{8m^{2}}D_{0}\Pi^{2}\psi_{s}-\dfrac{i}{4m^{2}}\Pi D_{0}\Pi \psi_{s}-\dfrac{1}{16m^{3}}\Pi^{4}\psi_{s}.
\label{equation 1}
\end{align}
This is the final form that can be compared to the Pauli-Schr{\"o}dinger equation to extract the physical quantities of interest.  Before doing that, 
we point out that this equation seems to mix different orders in the expansion in the inverse of the mass, but when incorporated the nonrelativistic dimensional setting through the rescaling $t\rightarrow mt$ and $A_0\rightarrow \frac{1}{m}A_0$, it becomes transparent the dominance order of the terms in (\ref{equation 1}),
\begin{align}
iD_{0}\psi_{s}-\dfrac{1}{2}\Pi^{2}\psi_{s}=\dfrac{i}{8m^{2}}D_{0}\Pi^{2}\psi_{s}-\dfrac{i}{4m^{2}}\Pi D_{0}\Pi \psi_{s}-\dfrac{1}{16m^{2}}\Pi^{4}\psi_{s}.
\label{equation 1a}
\end{align}
The left hand side corresponds to the marginal operators while the right hand side contains the leading corrections due to irrelevant operators.

\subsection{The $g$-factor of the Electron}

The $g$-factor of the electron is the proportionality constant between the magnetic moment $\vec{\mu}$ and the spin $\vec{S}$, namely, $\vec{\mu}=g \mu_0 \vec{S}$, where $\mu_0\equiv e/2m$ and $\vec{S}\equiv \vec{\sigma}/2$. It can be read out as we identify the following contribution in the Hamiltonian
\begin{equation}
-\vec{\mu}\cdot \vec{B}=-g\mu_0\vec{S}\cdot\vec{B}.
\label{3.1}
\end{equation} 
Such contribution comes from the lowest order, corresponding to the left hand side of (\ref{equation 1}), 
\begin{equation}
iD_{0}\psi_{s}-\dfrac{1}{2m}\Pi^{2}\psi_{s}= 0+\cdots.
\label{3.2}
\end{equation}
Remembering that we are in the static case, we can extract a phase $e^{-iEt}$ of the wave function and replace
$i D_{0}\rightarrow E-eV$, so that
\begin{align}
E\psi_{s}=H_0\psi_s\equiv\left(\dfrac{1}{2m}\Pi^{2}+eV \right)\psi_{s}. 
\label{hamiltoniana livre}
\end{align}
Now it is easy to identify the desired term
\begin{align}
E\psi_{s}&=\dfrac{1}{2m}\Pi^{2}\psi_{s}+eV\psi_{s} \nonumber\\
		&=\dfrac{1}{2m}\sigma^{i}\sigma^{j}(-i\nabla^{i}-eA^{i})(-i\nabla^{j}-eA^{j})\psi_s+eV\psi_s \nonumber \\
		&=\dfrac{1}{2m}(i\vec{\nabla}+e\vec{A})^{2}\psi_{s}+\dfrac{i}{2m} \sigma^{k}\epsilon^{ijk}(i\nabla^{i}+eA^{i})(i\nabla^{j}+eA^{j})\psi_{s}+eV\psi_{s}\nonumber\\
		&=\dfrac{1}{2m}(\vec{i\nabla}+e\vec{A})^{2}\psi_{s}-\dfrac{e}{2m} \vec{\sigma}\cdot \vec{B} \psi_{s}+eV \psi_{s}\nonumber\\
		&=\dfrac{1}{2m}(\vec{i\nabla}+e\vec{A})^{2}\psi_{s}- 2  \mu_0 \vec{S}\cdot \vec{B} \psi_{s}+eV \psi_{s}.
		\label{3.3}
\end{align}
Comparing this result with (\ref{3.1}) it follows immediately the remarkable Dirac's result $g=2$. 

It is interesting to discuss now another type of term that contributes to the $g$-factor of the electron. This is the so-called Pauli term and involves a nonminimal  coupling $F_{\mu\nu}\bar{\Psi}\sigma^{\mu\nu}\Psi$, with $F_{\mu\nu}=\partial_{\mu}A_{\nu}-\partial_{\nu}A_{\mu}$ and $\sigma_{\mu \nu}\equiv \dfrac{i}{2}\left[\gamma^{\mu},\gamma^{\nu}\right]$. In the relativistic dimensional setting this is an irrelevant operator 
$[F_{\mu\nu}\bar{\Psi}\sigma^{\mu\nu}\Psi]=5$. To be included in the Lagrangian it need to be compensated with a mass factor
\begin{align}
\mathcal{L}_{P}=-\dfrac{e\kappa}{8m}\bar{\Psi}\left(\sigma_{\mu \nu}F^{\mu \nu}\right)\Psi,
\label{3.4}
\end{align}
where $\kappa$ is an arbitrary numerical factor. With the inclusion of this term, the $g$-factor is shifted by $\kappa$, i.e., $g=2+\kappa$. At the first sight it is a little weird that such irrelevant term contributes to the $g$-factor as the marginal coupling $\bar{\Psi}\slashed{A}\Psi$ does. 
It is worth to repeat the procedure of integration of high-energy modes with the inclusion of the Pauli coupling to see the fate of each one of these operators in the low-energy effective action.

In terms of the Pauli matrices
\begin{align}
\sigma_{0k}=i \sigma^{3} \times \sigma^{k}=i \left(\begin{array}{cc}
 \sigma^{k} & 0 \\ 
0 & -\sigma^{k}
\end{array} \right) \;\;\;\;\;\; \text{and} \;\;\;\;\;\; \sigma_{ij}=\epsilon^{ijk}\sigma^{0}\times\sigma^{k}= \left(\begin{array}{cc}
\sigma^{k} & 0 \\ 
0 & \sigma^{k}
\end{array} \right).
\label{3.5}
\end{align}
With this we can write the Pauli coupling in terms of the Weyl spinors 
\begin{align}
\mathcal{L}_{P}=\dfrac{e\kappa}{4m}\left[i\left(\varphi_{-}^{\dagger}\vec{\sigma}\varphi_{+}-\varphi_{+}^{\dagger}\vec{\sigma}\varphi_{-}\right)\cdot \vec{E}+\left(\varphi_{-}^{\dagger}\vec{\sigma}\varphi_{+}+\varphi_{+}^{\dagger}\vec{\sigma}\varphi_{-}\right)\cdot\vec{B}\right].
\label{3.6}
\end{align}
In terms of the low and high-energy components $\psi_L$ and $\psi_H$, this becomes 
\begin{align}
\mathcal{L}_{P}=\dfrac{e\kappa}{4m}\left[i\left(\psi_{L}^{\dagger}\vec{\sigma}\psi_{H}-\psi_{H}^{\dagger}\vec{\sigma}\psi_{L}\right)\cdot \vec{E}+\left( \psi_{L}^{\dagger}\vec{\sigma}\psi_{L}-\psi_{H}^{\dagger}\vec{\sigma}\psi_{H}\right) \cdot \vec{B}\right].
\label{3.7}
\end{align} 
As we are interested in computing the correction to the electron's $g$-factor, we make things easier by taking $\vec{E}=0$. 

With the inclusion of the Pauli term, the Lagrangian (\ref{2.3}) gives place to
\begin{align}
\mathcal{L}={\psi}_{L}^{\dagger}\left(iD_{0} +\dfrac{e\kappa}{4m}\vec{\sigma}\cdot\vec{B} \right)\psi_L+{\psi}_{H}^{\dagger}\left(iD_{0}+2m - \dfrac{e\kappa}{4m}\vec{\sigma}\cdot\vec{B}\right) {\psi}_{H}-{\psi}_{L}^{\dagger}\Pi{\psi}_{H}-{\psi}_{H}^{\dagger}\Pi{\psi}_{L}.
\label{3.8}
\end{align}
We have to follow the same steps we did before, namely, integrating out $\psi_{H}$ to get the effective action for $\psi_{L}$ (which we called simply $\psi$). 
The difference is the operator $(iD_0+2m- \dfrac{e\kappa}{4m}\vec{\sigma}\cdot\vec{B})$ instead of $(iD_0+2m)$, which will modify the Green function  
(\ref{2.6}). However, as the $g$-factor comes from the lowest order in the expansion of the Green function such a difference will affect only higher orders. Thus the 
lowest-order contribution to the $g$-factor due to the Pauli coupling comes only from the first bracket in (\ref{3.8}), that is
\begin{align}
S_{eff}=\int d^3x dt\left[ {\psi}^{\dagger}\left(iD_{0}+\dfrac{e\kappa}{4m}\vec{\sigma}\cdot\vec{B} \right)\psi-\dfrac{1}{2m}\psi^{\dagger}\Pi^{2}\psi\right]+\cdots.
\label{3.9}
\end{align}
This effective action is useful in understanding the question of relevance of operators $\bar{\Psi}\slashed{A}\Psi$ and $F_{\mu\nu}\bar{\Psi}\sigma^{\mu\nu}\Psi$. We see that they play very different role in the process of integration of high-energy modes. While the lowest-order contribution due to the minimal coupling does participate of the integration, the lowest-order contribution due to the Pauli coupling is innocuous to it. The result is that both terms contribute in the same way in the nonrelativistic limit. In the effective field theory language, although the Pauli term is irrelevant in the relativistic regime it gives a leading contribution (marginal) in the nonrelativistic limit. This is naturally incorporated within the nonrelativistic dimensional setting discussed previously. Upon the rescaling $t\rightarrow mt$ and $A_0\rightarrow \frac{1}{m}A_0$ the effective action (\ref{3.9}) becomes
\begin{equation}
S_{eff}=\int d^3x dt\left[ {\psi}^{\dagger}\left(iD_{0}+\dfrac{e\kappa}{4}\vec{\sigma}\cdot\vec{B} \right)\psi-\dfrac{1}{2}\psi^{\dagger}\Pi^{2}\psi\right]+\cdots,
\label{3.10}
\end{equation}
which shows that the operator $\psi^{\dagger}\vec{B}\psi$ is marginal, i.e., $[\psi^{\dagger}\vec{B}\psi]=5$ (remember that $[\vec{B}]=2$), in the same way as the operator 
$\psi^{\dagger}\Pi^{2}\psi$.

Before close this section we just complete the computation of the shift in the $g$-factor. The equation of motion following from (\ref{3.9}) is
\begin{align}
iD_{0}\psi+\dfrac{e\kappa}{4m}\vec{\sigma} \cdot \vec{B}\psi =\dfrac{1}{2m}\Pi^{2}\psi +\cdots.
\label{3.11}
\end{align}
Proceeding with similar calculations that led to (\ref{3.3}), we easily obtain  
\begin{align}
E\psi &=\dfrac{1}{2m}\Pi^{2}\psi +eV\psi- \kappa\mu_0\vec{S}\cdot\vec{B}\psi \nonumber \\
		&=\dfrac{1}{2m}\left(i\vec{\nabla}+e\vec{A}\right)^{2}\psi-(2+\kappa)\mu_0 \vec{S}\cdot \vec{B}\psi+eV\psi,
		\label{3.12}
\end{align}
yielding to the result $g=2+\kappa$.

\subsection{Fine Structure}

The subleading contributions due to the irrelevant operators in the right hand side of (\ref{equation 1}) gives rise to the fine structure. They can be treated as a perturbation $H_{1}$ on the Hamiltonian $H_0$ defined in (\ref{hamiltoniana livre}). As we did before, in the static case we firstly extract the time-dependent phase of the wave function, $i D_{0}\psi_s\rightarrow (E-eV)\psi_s$. Next, we can take advantage that in $H_{1}\psi_s$, we can replace
\begin{align}
 (E-eV)\psi_s \rightarrow \dfrac{1}{2m}\Pi^{2}\psi_s+\cdots,
\label{4.1}
\end{align}
since the difference will affect only higher order terms. By using this in the right hand side of (\ref{equation 1}), we can read out the Hamiltonian $H_1$,
\begin{align}
H_1 &=-\dfrac{e}{8m^{2}}\left(\Pi^{2}V+V\Pi^{2}-2\Pi V \Pi\right)-\dfrac{1}{8m^{3}}\Pi^{4}\nonumber\\
&=\dfrac{e}{4m^{2}}\left(\Pi\left[V,\Pi\right]-\frac{1}{2}\left[V,\Pi^{2}\right]\right)-\dfrac{1}{8m^{3}}\Pi^{4}\nonumber \\
&=\dfrac{e}{4m^{2}}\left(\frac{1}{2}\Pi\left[V,\Pi\right]-\frac{1}{2}\left[V,\Pi\right]\Pi\right)-\dfrac{1}{8m^{3}}\Pi^{4}\nonumber\\
&=\dfrac{e}{8m^{2}}\left[\Pi,\left[V,\Pi\right]\right]-\dfrac{1}{8m^{3}}\Pi^{4}.
\label{4.2}
\end{align}
With a little more work we can put it in a familiar form. By using that $[V,\Pi]=-i \vec{\sigma}\cdot \vec{E}$, it becomes
\begin{align}
H_1=-\dfrac{e}{8m^{2}}\left(\vec{\nabla}\cdot\vec{E}+2 \vec{\sigma}\cdot\left( \vec{E} \times\vec{P}\right) \right) -\dfrac{1}{8m^{3}}\Pi^{4},
\label{4.3}
\end{align}
where we are using the suggestive notation $\vec{P}\equiv i\vec{D}$. With this expression is immediate to get the final form. In the first term of the bracket we employ the Gauss's law $\vec{\nabla}\cdot \vec{E}=\rho=e \delta^{3}(\vec{r})$, which leads us to the familiar Darwin correction. By considering the case of a central electrostatic field ($\vec{A}=0$), the covariant derivative reduces to the ordinary one, i.e., $\vec{P}\rightarrow -i\vec{\nabla}\equiv\vec{p}$, and the second term of the bracket becomes
\begin{align}
\vec{E}\times\vec{p}=-\dfrac{dV}{dr}\hat{r}\times\vec{p}=-\dfrac{1}{r}\dfrac{dV}{dr}\vec{L},
\label{4.4}
\end{align}
which is the usual spin-orbit coupling. The last term of (\ref{4.3}) is promptly identified as the relativistic correction. Collecting all contributions, it follows the fine structure Hamiltonian
\begin{align}
H_1=-\dfrac{e^{2}}{8m^{2}} \delta^{(3)}(\vec{r})+\dfrac{e}{4m^{2}}\dfrac{1}{r}\dfrac{dV}{dr}\vec{\sigma}\cdot \vec{L}-\dfrac{1}{8m^{3}} |\vec{p}|^{4}.
\label{4.5}
\end{align}


\section{Final Remarks}\label{S4}

In this work we have analyzed the low-energy nonrelativistic regime of the Dirac theory in the framework of the effective field theory. We start by defining the partition function of the Dirac field coupled to a gauge field and proceed by integrating out the high-energy modes of the Dirac spinor, given in terms of a combination of the Weyl spinors. We then obtain an effective action for the remaining components which gives rise to an equation of motion that can be compared to the Pauli-Schr{\"o}dinger one. To properly accomplish this, it is necessary to manually implement the normalization condition of the wave function. In this respect it would be interesting to implement a Foldy-Wouthuysen-like transformation in the partition function and then integrating out the high-energy energy components. It is expected that in this case the resulting effective action will furnish an equation of motion whose wave function needs not to be amend by the requirement of unitarity.

Concerning with the low-energy effective action, we classify leading and sub-leading operators within the anisotropic dimensional analysis which is suitable for nonrelativistic theories. It involves a rescaling of the time coordinate as well as other time-like quantities (as the zero component of the gauge field) in order to enforce the leading time derivatives to be on an equal foot to the leading spatial derivatives. This analysis enables us to understand how some naively irrelevant operators in the relativistic dimensional setting are actually marginal and give leading contributions to the nonrelativistic theory.

This type of dimensional analysis is very common in condensed matter systems where Lorentz symmetry is generally absent \cite{Hornreich,Vojta}. In recent years it has attracted a lot of attention of high-energy community due to the possibility of being implemented in the gravitational context, producing a power-counting renormalizable quantum theory of gravity at the price of Lorentz invariance \cite{Horava2}. In this scenario Lorentz symmetry was expected to be recovered in the low-energy limit. 
 
We conclude by pointing out the our whole analysis provides an interesting perspective on the question of the nonrelativistic regime of Dirac theory, placing it in a more general context that represents the current understanding of quantum field theories.


\section{Acknowledgments}

We would like to thank Marcelo Gomes and Carlos Hernaski for very helpful discussions.
We acknowledge the financial support of Brazilian agencies CAPES and CNPq.


\appendix

\section{Fermionic Path Integral} \label{AA}

In this appendix we work out some details of the calculations involved in computing the effective action in (\ref{2.5}). We have essentially to deal with the following Gaussian fermionic path integral:
\begin{equation}
e^{iS_{eff}[\psi_L,A_0,\vec{A}]}\equiv \int \mathcal{D}\psi_{H}\mathcal{D}\psi_{H}^{\dagger}  \exp \;i \int d^{4}x \left [ \psi^{\dagger}_{L}iD_{0}\psi_{L}  +\psi_{H}^{\dagger}\left(iD_{0}+2m\right) {\psi}_H-{\psi}_L^{\dagger}\Pi{\psi}_H-{\psi}_H^{\dagger}\Pi{\psi_L} \right].
\label{aa1}
\end{equation}
We see that the low and high-energy spinors are coupled. Thus we shift the integration variable, $\psi_{H} \rightarrow \psi_{H} +\int d^{4}y G(x,y) \Pi \psi_{L}(y)$, with a corresponding shift for $\psi_H^{\dagger}$. The object $G(x,y)$ is defined as $G(x,y)\equiv \delta^{(3)}(\vec{x}-\vec{y})G_0(x,y)$, with $G_0(x,y)$ being the Green function of the operator $iD_{0}+2m$, 
\begin{equation}
(iD_{0}+2m)G_0(x,y)=\delta(x^0-y^0).
\label{aa1a}
\end{equation}
Proceeding with the derivation, we obtain
\begin{eqnarray}
e^{iS_{eff}[\psi_L,A_0,\vec{A}]}&=&\exp i \left[\int d^4x\psi^{\dagger}_{L}iD_{0}\psi_{L}- \int d^{4}x  d^{4}y \psi_{L}^{\dagger}(y) \Pi (y) G(x,y)\Pi (x)\psi_{L}(x)  \right]\nonumber\\ 
&\times&\int \mathcal{D}\psi_{H}\mathcal{D}\psi_{H}^{\dagger}\exp i \int d^4x  \psi_{H}^{\dagger}\left(iD_{0}+2m\right) {\psi}_H.
\label{aa2}
\end{eqnarray}
With the above changes we have decoupled the components $\psi_L$ and $\psi_H$. Even without knowing the result of the functional integration over $\psi_H$ and $\psi_H^{\dagger}$, the important point is that it does not depend on $\psi_L$ and $\psi_L^{\dagger}$. Consequently, this will produce an additive factor independent of $\psi_L$ and $\psi_L^{\dagger}$ in the effective action. Therefore it will not contribute in the subsequent calculations in the body of the manuscript that make use of equation of motion of $\psi_L$. Even so, for completeness we briefly discuss some basic facts about Grassmann variables. At the same time we invite the reader to consult, for example, Chap. 14 of the Ref. \cite{Schwartz}. 

Before doing that we just give the result of the functional integration over fermionic fields in the second line of (\ref{aa2}): 
\begin{equation}
\int \mathcal{D}\psi_{H}\mathcal{D}\psi_{H}^{\dagger}\exp i \int d^4x  \psi_{H}^{\dagger}\left(iD_{0}+2m\right) {\psi}_H = \det \left(iD_{0}+2m\right). 
\label{aa3}
\end{equation}
The determinant can be exponentiated with help of the identity $\det A = \exp \text{Tr}\text{ln}A$. Using this in (\ref{aa2}) it follows immediately the effective action $S_{eff}$ in (\ref{2.5}).

Consider a single Grassmann variable $\theta$, i.e., a variable that satisfies $\theta^2=0$. An arbitrary function $f(\theta)$ can be expanded in powers of $\theta$ and will have the form $f(\theta)=a+b\theta$, with $a$ and $b$ being ordinary numbers. In particular, $e^{a\theta}=1+a\theta$. Next consider two independent Grassmann variables $\theta$ and $\bar{\theta}$, that in addition to $\theta^2=\bar{\theta}^2=0$ satisfy $\{\theta,\bar{\theta}\}=0 $. An arbitrary function will have the expansion $f(\theta,\bar{\theta})=a+b\theta +c \bar{\theta}+d \bar{\theta}\theta$. The exponential function in this case is $e^{a\bar{\theta}\theta}=1+a\bar{\theta}\theta$.
 
To proceed we introduce the operations of integration and derivation of Grassmann variables. The basic definitions are
\begin{equation}
 \int d\theta (1)\equiv \frac{\partial }{\partial \theta} (1)\equiv 0~~~ \text{and}~~~ \int d\theta \theta \equiv \frac{\partial }{\partial \theta} \theta\equiv 1.
\end{equation}
This implies, in particular, that
\begin{equation}
\int d \theta d\bar{\theta} e^{a \bar{\theta}\theta} = a. 
\end{equation}
The generalization of this integral for the case of an arbitrary number of Grassmann variables $\theta_i$ and $\bar{\theta}_i$, $i=1,...,n$, leads to the desired result, 
\begin{equation}
\int \prod_i d\theta_i d\bar{\theta}_i  e^{\sum_{i,j}\bar{\theta}_i A_{ij}\theta_j}=\det A.
\end{equation}
The functional integration can be transformed into an ordinary integral of this type simply by discretizing the spacetime coordinates, i.e., defining the system over a lattice whose site are specified by the indexes $i$ and $j$.


\section{Green Function}\label{AB}

This appendix is dedicated to the computation of the Green function for the operator $(iD_{0}+2m)$. We will consider the static case where the gauge field $A_{\mu}$ does not depend on the time. First we write $G_0(x,y)$ in terms of the Fourier transform,
\begin{equation}
G_0(x,y)=\int \frac{dk_0}{2\pi}e^{-ik_0(x_0-y_0)}G(k_0,\vec{x}). 
\end{equation} 
Remember that $G_0(x,y)$ is always multiplied by $\delta^{(3)}(\vec{x}-\vec{y})$ to compose $G(x,y)$, so that $G(k_0,\vec{x})$ depends only on a spatial coordinate 
$\vec{x}$. The equation (\ref{2.5a}) then implies 
\begin{equation}
G(k_0,\vec{x})=\frac{1}{k_0-eA_0(\vec{x})+2m},
\end{equation}
such that,
\begin{equation}
G_0(x,y)=\int \frac{dk_0}{2\pi}e^{-ik_0(x_0-y_0)}\frac{1}{k_0-eA_0(\vec{x})+2m}.
\end{equation}
Now we can perform the expansion for low energies compared to the mass scale $m$ and weak field, 
\begin{eqnarray}
G_0(x,y)&=&\frac{1}{2m}\int \frac{dk_0}{2\pi}e^{-ik_0(x_0-y_0)} \frac{1}{1+\frac{k_0-eA_0}{2m}}\nonumber\\
&=&\frac{1}{2m}\int \frac{dk_0}{2\pi}e^{-ik_0(x_0-y_0)}\sum_{n=0}^{\infty}(-1)^n \left(\frac{k_0-eA_0}{2m}\right)^n.
\end{eqnarray}
By noting that $f(k_0)e^{-ik_0(x_0-y_0)}=f(i\partial_{x_0})e^{-ik_0(x_0-y_0)}$, this expression can be written  in terms of derivatives of the delta function, i.e., 
\begin{equation}
G_0(x,y)=\frac{1}{2m}\sum_{n=0}^{\infty}(-1)^n \left(\frac{i\partial_{x_0}-eA_0}{2m}\right)^n\delta(x_0-y_0).
\end{equation}
Combining with the spatial delta $\delta^{(3)}(\vec{x}-\vec{y})$ to form $G(x,y)$, it follows the result showed in (\ref{2.6}).


\section{U(1) Noether Current} \label{AC}

The Noether theorem deeply connects symmetries to conservation laws. We will discuss here a simple way to construct this relationship and then we apply it to construct the $U(1)$ current of the Dirac theory. 

Consider a relativistic action $S=\int d^Dx \mathcal{L}[\phi]$, where $\phi$ is a generic field. Assume that under an infinitesimal transformation of the field, $\phi\rightarrow \phi+\epsilon\delta\phi$, with $\epsilon$ a constant (global), the action is invariant. This corresponds to a symmetry. To find the conserved current we promote the global parameter $\epsilon$ to a local one, $\epsilon\rightarrow\epsilon(x)$. In this case, the transformation $\phi\rightarrow \phi+\epsilon(x)\delta\phi$ is no longer a symmetry. The variation of the action must involve the derivative of the parameter, $\partial_{\mu}\epsilon(x)$, which recovers the invariance for the global transformation. Lorentz invariance implies the existence of a current $j^{\mu}$, such that $\delta S=\int d^Dx j^{\mu}\partial_{\mu}\epsilon(x)$.

The local transformation $\phi\rightarrow \phi +\epsilon(x)\delta\phi$, on the other hand, can be considered as an arbitrary variation of the field, in which case gives the equation of motion by imposing that the action is stationary under this transformation. Therefore, if we use the equations of motion, we will have
\begin{equation}
\delta S=\int d^Dx j^{\mu}\partial_{\mu}\epsilon(x)=0~(\text{eq. of motion}).
\end{equation}
With an integration by parts and using the fact that $\epsilon(x)$ is arbitrary, we conclude that the current $j^{\mu}$ is conserved,  
\begin{equation}
\partial_{\mu}j^{\mu}=0~~~\Rightarrow~~~\frac{d}{dt}\int d^{D-1}x j_0=0.
\end{equation}

Now we can easily apply this procedure to the case of the Dirac action
\begin{align}
S=\int d^{4}x \; \bar{\Psi}\left(i \slashed{\partial} -m\right)\Psi.
\end{align}
It is invariant under the $U(1)$ transformations $\Psi\rightarrow\Psi^{\prime}=e^{i\epsilon}\Psi $ and $\bar{\Psi}\rightarrow\bar{\Psi}^{\prime}=e^{-i\epsilon}\bar{\Psi}$, whose infinitesimal forms are
\begin{equation}
\delta \Psi=i \epsilon \Psi ~~~ \text{and} ~~~\delta \bar{\Psi}=-i\epsilon\bar{\Psi}.
\end{equation}
By making $\epsilon\rightarrow\epsilon(x)$ and computing the variation of the action in this situation, we obtain 
\begin{align}
\delta S&=\int d^{4}x \; i \left( \delta \bar{\Psi}\slashed{\partial}\Psi+\bar{\Psi}\slashed{\partial}\delta\Psi	\right)\nonumber \\
		&=\int d^{4}x \;i\left( -i\epsilon \bar{\Psi} \slashed{\partial}\Psi+i\epsilon \bar{\Psi}\slashed{\partial}\Psi +i\bar{\Psi}\gamma^{\mu}\Psi \partial_{\mu}\epsilon\right)\nonumber \\
		&= \int d^{4}x \; \epsilon \partial_{\mu}\left(\bar{\Psi} \gamma^{\mu}\Psi \right),
\end{align}
from which it follows immediately the $U(1)$ current
\begin{align}
j^{\mu}=\bar{\Psi}\gamma^{\mu}\Psi.
\end{align}


\end{document}